\documentclass[a4paper]{jpconf}
\usepackage{graphicx}
\usepackage{amsmath}
\usepackage{lineno}
\usepackage{mathrsfs} 
\usepackage{amssymb}
\usepackage{amsfonts}
\usepackage{graphicx}
\usepackage{url}
\usepackage{array}
\usepackage{verbatim}
\usepackage{subfigure}
\usepackage{listings}
\usepackage{color}
\usepackage{lscape}
\usepackage{setspace}
\usepackage{amssymb}
\usepackage[latin1]{inputenc}

\renewcommand\vec[1]{\ensuremath\boldsymbol{#1}}

\begin{document}

\title{A proposed search for new light bosons using a table-top neutron Ramsey apparatus}

\author{F.M. Piegsa$^1$, G. Pignol$^2$}
\address{$^1$ ETH Z\"urich, Institute for Particle Physics, 8093 Z\"urich, Switzerland}
\address{$^2$ LPSC, Universit\'e Joseph Fourier Grenoble 1, CNRS/IN2P3, INPG, 38000 Grenoble, France}
\ead{florian.piegsa@phys.ethz.ch}

\begin{abstract}
If a new light boson existed, it would mediate a new force between ordinary fermions, like neutrons. In general such a new force is described by the Compton wavelength $\lambda_{\text{c}}$ of the associated boson and a set of dimensionless coupling constants. For light boson masses of about 10$^{-4}$ eV/c$^2$, $\lambda_{\text{c}}$ is of the order millimeters. Here, we propose a table-top particle physics experiment which provides the possibility to set limits on the strength of the coupling constants of light bosons with spin-velocity coupling. It utilises Ramsey's technique of separated oscillating fields to measure the pseudo-magnetic effect on neutron spins passing by a massive sample. 


\end{abstract}

\section{Introduction}

In particle physics, the interaction between matter particles (fermions) is understood as an exchange of vector particles (bosons). The Standard Model of particle physics explains electromagnetism, nuclear weak and strong interactions in terms of exchange of photons, weak heavy bosons and gluons. It is believed that the Standard Model is a low energy limit of some more fundamental theory, that will ultimately also describe gravity in the same framework. Various extensions of the Standard Model based on new ideas, such as Supersymmetry or Superstring theory are being developed in this direction. As a general feature of such theoretical investigations, new particles, and thus new interactions beyond the standard model, are expected. These hypothetical new particles have escaped detection up to now, either because they are too heavy, thus mediating an interaction which is too short in range, or because they are too weakly coupled with ordinary matter, thus mediating an interaction which is too weak. The latter case is of interest for this proposal: we are looking for a weak, long range new interaction.
This new interaction between ordinary fermions would be mediated by a new light boson with an interaction range $\lambda_{\text{c}} = \hbar / M c$, where $M$ is the mass of the new boson, $\hbar$ is Planck's constant and $c$ is the speed of light. For a light mass of $10^{-4}$ eV/c$^2$, the range is about 2 mm. Depending on the details of the fermion-boson coupling and whether the new boson is scalar (spin 0) or vectorial (spin 1), such a force between two fermions could depend on the spins of the fermions or not. It could also depend on the relative velocity between the fermions. For a scalar boson $\phi$ the interaction with a fermion $\psi$, described by the Lagrangian function $\mathcal{L} = \overline{\psi} (g_{\text{S}} + i g_{\text{P}} \gamma^5) \psi \phi$ is parametrised by two constants: the scalar coupling $g_{\text{S}}$ and the pseudoscalar coupling $g_{\text{P}}$. For a vector boson $X_\mu$ the interaction $\mathcal{L} = \overline{\psi} (g_{\text{V}} \gamma^\mu + g_{\text{A}} \gamma^\mu \gamma^5) \psi X_\mu$ is parametrised by the vector coupling $g_{\text{V}}$ and the axialvector coupling $g_{\text{A}}$. Hence, in general a new force is described by the range $\lambda_{\text{c}}$ and a set of dimensionless coupling constants.\\
Most current laboratory experiments regarding new light bosons do not search for the boson itself, but for the additional interaction the boson would mediate between ordinary particles. We restrict our discussion to the coupling with nucleons, although sensitive experiments probing the coupling with electrons have also been performed. The non-relativistic interaction potential between two fermions contains several terms, bearing diverse phenomenological consequences \cite{1}, calling for diverse experimental techniques to probe them. 
In the case of a spin 0 boson, several experiments investigate the gravity-like spin-independent scalar-scalar coupling, i.e. the so-called 5$^{\text{th}}$ force \cite{2,3,4,5,6}. Furthermore, the scalar-pseudoscalar coupling also called axion-like interaction is studied by measuring the influence of an unpolarised source on a polarised probe, e.g. on neutron spins or the nuclear spins of a polarised $^3$He gas etc. \cite{7,8,9,10}. Finally, for experiments investigating the pseudoscalar-pseudoscalar coupling the source as well as the probe have to be polarised. A recent experiment performed at Princeton probed this coupling at a meter distance, using a $^3$He spin source and a K-$^3$He co-magnetometer probe \cite{12}.
Similarly, a spin 1 boson can mediate vector-vector, vector-axial and axial-axial interactions. The first type corresponds to a spin-independent repulsive new force, which is probed in the millimeter region by the Seattle torsion pendulum experiment \cite{4}. The most recent measurement yields the upper limit:
\begin{equation}
g_{\text{V}}^2 < 5 \times 10^{-40}
\end{equation}
Next, there are terms depending on the spins of the two interacting particles (spin-spin interaction). Ref. \cite{12} reports the limits, in terms of vector exchange parameters: 
\begin{equation}
g_{\text{V}} g_{\text{A}}  \left( 1 + \frac{r}{\lambda_{\text{c}}} \right) e^{-r/\lambda_{\text{c}}} < 5 \times 10^{-25}
\end{equation}
and
\begin{equation}
g_{\text{A}}^2 e^{-r/\lambda_{\text{c}}} < 1.5 \times 10^{-40}
\end{equation}
where in this experiment $r = 0.5$ m is the distance from the $^3$He spin source with the K-$^3$He co-magnetometer. In addition to the spin-spin interaction, a vector boson would mediate two kinds of spin-velocity interactions between a source particle and a probe particle: 
\begin{equation}
V_{\text{VA}}^{\rm{point}}(r) = \frac{g_{\text{V}} g_{\text{A}}}{2 \pi} \ \hbar c \ \vec{\sigma} \cdot \frac{\vec{v}}{c} \ \frac{e^{-r/\lambda_{\text{c}}}}{r}
\end{equation}
and
\begin{equation}
V_{\text{AA}}^{\rm{point}}(r) = 
\frac{g_{\text{A}}^2}{16 \pi} \ \frac{(\hbar c)^2}{m c^2} \ \vec{\sigma} \cdot \left( \frac{\vec{v}}{c} \times \frac{\vec{r}}{r} \right) \ \left( \frac{1}{\lambda_{\text{c}}} + \frac{1}{r}  \right) \  \frac{e^{-r/\lambda_{\text{c}}}}{r}
\end{equation}
where $\vec{\sigma}$ is the spin and $m$ is the mass of the probe particle, $r$ is the distance between the source and the probe and $v$ is the relative velocity between the source and the probe. These potentials corresponds to $f_v$ and $f_{\bot}$ in the notations of \cite{1}.The fact that the interaction does not depend on the source particle's spin allows to use an unpolarised macroscopic body as a source, the potential of each particles in the body adding coherently. \\
Here, we propose to probe these exotic interactions, in the millimeter range, with a beam of polarised slow neutrons as a probe and macroscopic bulk matter as a source. Since the new interaction potentials are linear proportional to the neutron spin, it is equivalent to a interaction with a magnetic field which can be described by the relation $V = - \frac{1}{2}\gamma_{\text{n}} \hbar \ \vec{\sigma} \cdot \vec{B}$, where $\gamma_{\text{n}}$ is the gyromagnetic ratio of the neutron. Hence, in case of a semi-infinite plane of bulk material as a source, one can perform the integration over the source volume analytically to obtain the corresponding \emph{pseudo-magnetic fields}: 
\begin{equation}
B_{\text{VA}}(\Delta y) = \frac{2}{\gamma_{\text{n}}} g_{\text{V}} g_{\text{A}} \ N \lambda_{\text{c}}^2 \ \vec{v} \ e^{-\Delta y/\lambda_{\text{c}}}
\label{equ:B_VA}
\end{equation}
\begin{equation}
B_{\text{AA}}(\Delta y) = \frac{1}{\gamma_{\text{n}}} \frac{g_{\text{A}}^2}{4} \ N \ \frac{\hbar c}{m_{\text{n}} c^2} \ \lambda_{\text{c}} \ \left( \vec{v} \times \vec{e}_y \right) \ e^{-\Delta y/\lambda_{\text{c}}}
\label{equ:B_AA}
\end{equation}
where $\Delta y > 0$ is the distance from the probe particle, the neutron, to the surface, $N$ is the nucleon number density of the source and $m_{\text{n}}$ is the neutron mass. For the vector-axial case, the pseudo-magnetic field is directed along the neutron beam, whereas for the axial-axial case, the direction is orthogonal to both the neutron beam and the plane normal. These pseudo-magnetic fields induce corresponding pseudo-magnetic neutron precessions $\varphi_{\text{VA}} = -\gamma_{\text{n}} B_{\text{VA}}\tau$ and $\varphi_{\text{AA}} = -\gamma_{\text{n}} B_{\text{AA}} \tau$, where $\tau$ is the time the neutron spin experiences the pseudo-magnetic potential. A method to measure these precession angles very accurately is Ramsey's technique of separated oscillating fields adapted to neutrons \cite{13}.




\section{Proposed experimental setup}

\begin{figure}
	\centering
		\includegraphics[width=0.80\textwidth]{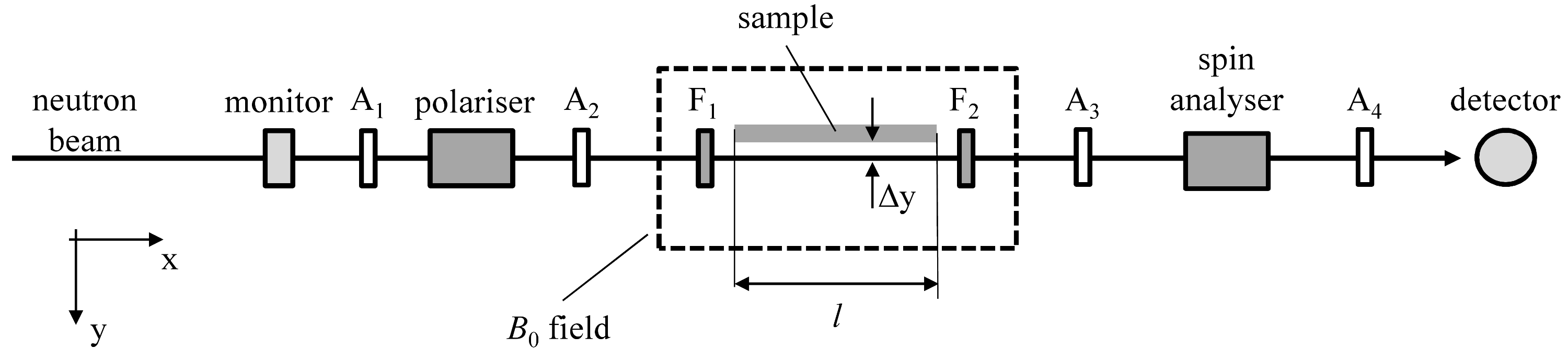}
	\caption{Topview of the proposed setup for the measurement of the axial-axial interaction between polarised neutrons and a macroscopic sample. The cold neutron beam is shaped by several apertures A$_{1-4}$. The sample is placed parallel in a distance $\Delta y$ with respect to the beam and is located between the two $\pi/2$-spin flippers F$_1$ and F$_2$. They produce a radio frequency field perpendicular to the static magnetic field $B_0$ which points into $z$-direction, i.e. perpendicular to the drawing plane.}
	\label{fig:setup}
\end{figure}

In Fig. \ref{fig:setup} a schematic drawing of the proposed Ramsey setup is depicted, which can be used to probe the axial-axial pseudo-magnetic field $B_{\text{AA}}$ as a function of the distance $\Delta y$ between the neutron beam and the sample surface. Firstly, a cold monochromatic neutron beam is polarised by means of a polarising supermirror in $z$-direction and collimated by several apertures to a width, which is much smaller than the beam-sample distance.\footnote{The neutron wavelength distribution should have a width of less than 10\% to avoid smearing of the Ramsey pattern due to magnetic field inhomogeneities of $B_0$.} Along the beam path two phase-locked radio frequency $\pi/2$-spin flippers are placed in a magnet which provides a static field $B_0$ in $z$-direction. The investigated sample with a length $l$ is placed between these $\pi/2$-spin flippers, which produce linear oscillating fields in $y$-direction. The spins of the neutrons are analysed with a second polarising supermirror and are finally detected employing a $^3$He gas detector. The number of detected neutrons is normalised using a monitor detector placed in the incidenting neutron beam. The additional appertures behind the sample (A$_3$ and A$_4$) avoid detecting neutrons which are accidentially scattered in air or on the sample surface.
\\
A so-called Ramsey oscillation pattern is obtained by measuring the detector count rate as a function of the spin flipper frequency close to the neutron Larmor resonance. Any additional phase precession due to the sample will cause a corresponding phase-shift of the sinusoidal-shaped Ramsey pattern. In order to avoid false effects the sample must be chosen to be non-magnetic and any phase drifts, e.g. caused by thermal expansion, phase drifts of the radio frequency fields, instabilities of the external magnetic field etc. must be suppressed as much as possible. Besides monitoring and stabilising the magnetic field, the implementation of a second reference neutron beam would be advantageous to correct for occurring phase deviations \cite{14,15}. In similar neutron Ramsey apparatuses a phase stability of better than 1° could be established \cite{15,16,17}. \\
Such a Ramsey setup could be easily installed at an already existing polarised neutron reflectometer, which would have the advantage of usually very precise alignment possibilities at such beam lines. A time-of-flight reflectometer would even allow measuring the velocity dependence of the exotic interaction for different neutron de-Broglie wavelengths. Moreover, a similar setup with a longitudinal static magnetic field, or equally a neutron spin-echo setup, can be used to investigate the vector-axial coupling.\\

\section{Sensitivity of the neutron Ramsey technique}

\begin{figure}
\begin{center}$
\begin{array}{cc}
\includegraphics[width=0.45\textwidth]{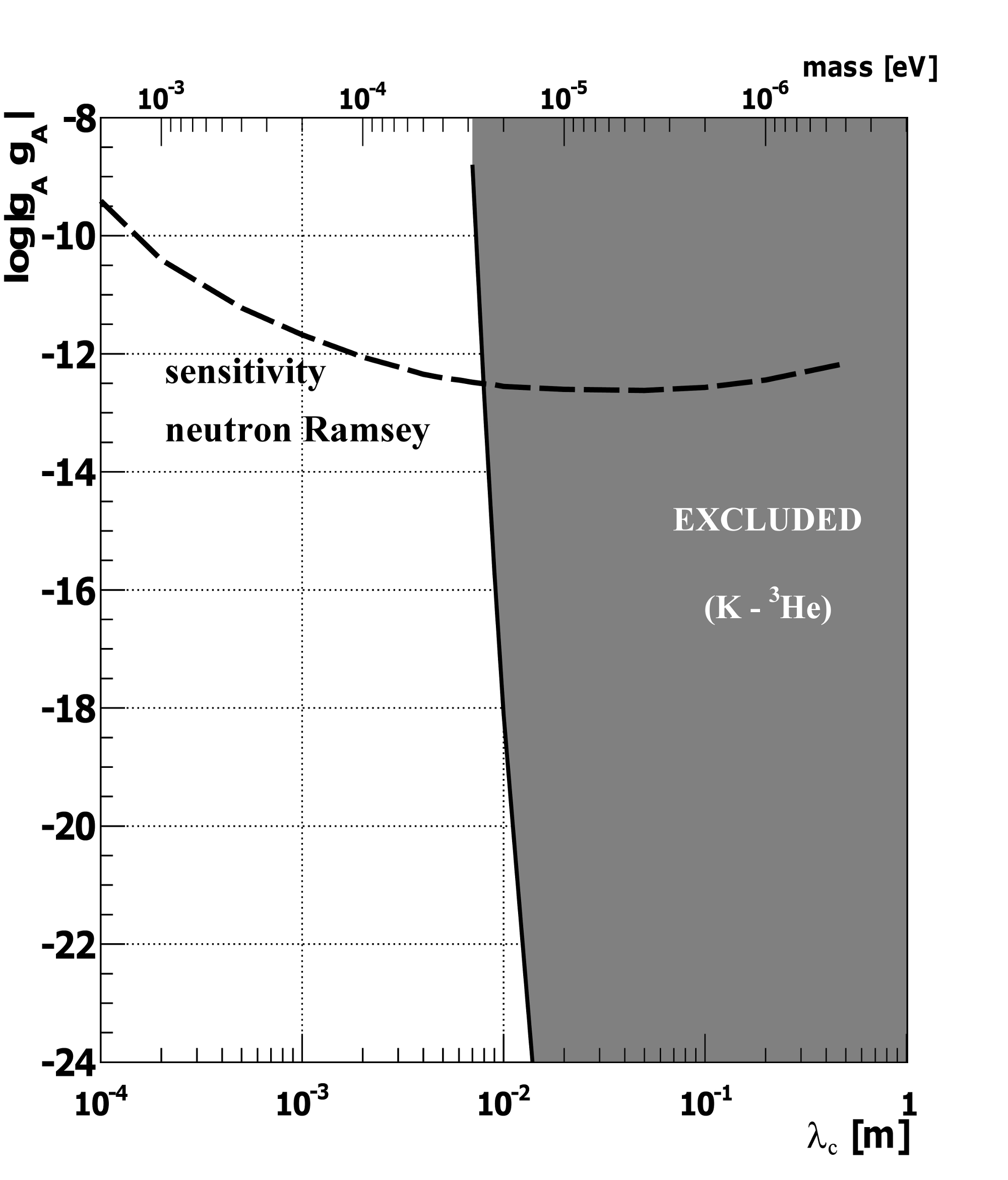} &
\includegraphics[width=0.45\textwidth]{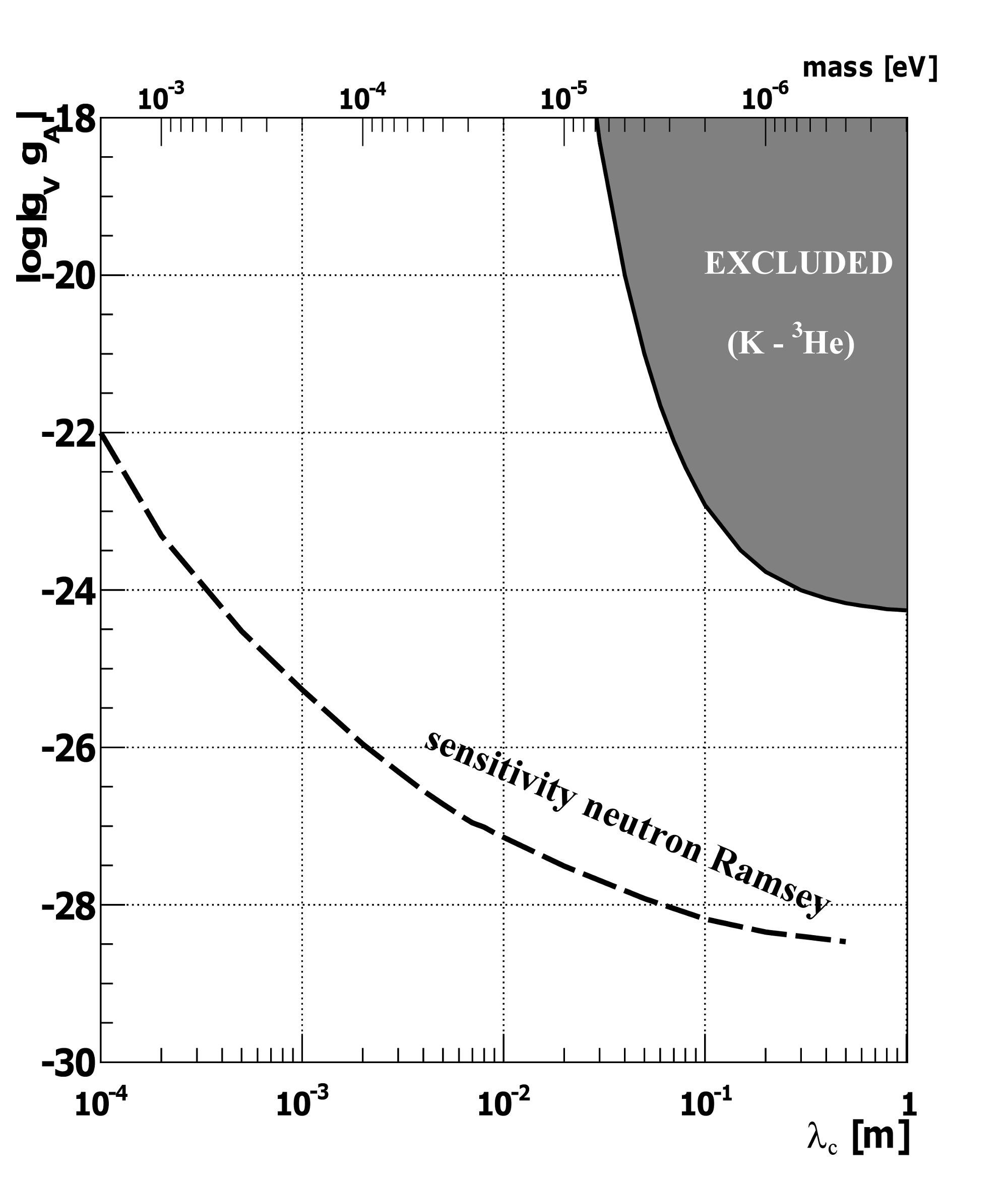}
\end{array}$
\end{center}
\caption{Expected sensitivity of the neutron Ramsey setup (dashed lines) for the axial-axial interaction (left) and the vector-axial interaction (right) as a function of the Compton wavelength $\lambda_{\text{c}}$. The grey shaded areas represent the exclusion by the Princeton experiment \cite{12}.}
\label{fig:vectoraxial}
\end{figure}
For the consideration of the achievable sensitivity to probe for the exotic interactions, we assume a phase stability of the setup of $\delta\varphi=1$°, a neutron wavelength of 0.5 nm, i.e. a velocity of $v\approx800$ m/s, and a 500 mm long aluminium sample with a plane surface. In order to minimise systematic errors due to the paramagnetic property of the aluminium ($\chi_{\text{para}}=2.1\times 10^{-5}$) the applied field $B_0$ should not exceed a few millitesla. For example, a field of 3 mT corresponds to a neutron Larmor resonance frequency of about 87.5 kHz and would lead to a paramagnetic phase-shift for a neutron beam crossing the entire sample, which corresponds to the worst case, of about +0.4° \cite{18}. Hence, applying the following formula:
\begin{equation}
\delta B  =   \frac{v}{\gamma_{\text{n}} \ l} \ \delta \varphi
\label{equ:magninter}
\end{equation}
one finds that the sensitivity of such an apparatus would be about 150 nT. Measuring with this sensitivity at a beam-sample distance $\Delta y=0.3$ mm, yields with  Eqn. (\ref{equ:B_VA}) and (\ref{equ:B_AA}) the exclusion plots for the axial-axial and vector-axial interactions presented in Fig. \ref{fig:vectoraxial}. Note, using a copper sample instead of aluminium would even increase the sensitivity by a factor of about 3, due to the higher density of the material.


\section{Conclusion}

We have proposed a table-top neutron Ramsey experiment to probe the strength of the coupling constants of light spin 1 bosons with spin-velocity coupling. The method has a sensitivity of about $10^{-11}$ for the axial-axial interaction and $10^{-25}$ for the vector-axial interaction at $\lambda_{\text{c}}=1$ mm and represents, to our knowledge, so far the only method to probe spin-velocity interactions in the millimeter range.

\end{document}